\documentclass[12pt]{article}
\usepackage{amsfonts}
\usepackage{amsmath}
\usepackage{amssymb}
\usepackage{amsthm}
\usepackage{bigints}
\usepackage[font=small,labelfont=bf]{caption}
\usepackage{color}
\usepackage{csquotes}
\usepackage{float}
\usepackage{footmisc}
\usepackage{graphicx}
\usepackage{epstopdf}
\usepackage{geometry}
\usepackage{hyperref}
\usepackage{hypcap}
\usepackage{mathtools}
\usepackage{natbib}
\usepackage{pdfpages}
\usepackage{pgfplots}
\usepackage{rotating}
\usepackage{setspace}
\usepackage{subfig}
\usepackage{tikz}
\usetikzlibrary{calc}
\usetikzlibrary{patterns}
\usepackage{xcolor}

\usetikzlibrary{calc}

\newcommand*{\Resize}
[2]{\resizebox{\textwidth}{!}{$#2$}}
\newcommand\myeq{\mathrel{\stackrel{\makebox[0pt]{\mbox{\normalfont\tiny
l'Hopital}}}{=}}}

\newtheorem{corollary}{Corollary}

\newtheorem{lemma}{Lemma}

\newtheorem{proposition}{Proposition}

\hypersetup{
colorlinks,
linkcolor={red!50!black},
citecolor={blue!50!black},
urlcolor={blue!80!black}
}

\begin{document}

\title{Information Asymmetry and Search Intensity%
	\thanks{{\small 
		{I would like to thank M. Janssen, A. Parakhonyak, and 
		A. Sobolev for helpful suggestions and discussion.This 
		is a first draft that may contain mistakes and 
		missing references. Its aim is to share preliminary 
		results of the research project so that I can gather 
		information from fellow researchers their opinion on the 
		project and literature they can recommend. Thus, please, 
		send me your suggestions and comments.
		}}}}
\author{Atabek Atayev\thanks{{\small {Market Design, 
ZEW---Leibniz Centre for European Economic Research, L 7, 1,	
68161 Mannheim, Germany. E.: atabek.atayev@zew.de }}}}
\date{\today\\
}

\begin{singlespace}
	\maketitle
\end{singlespace}
\begin{abstract}
	\noindent 
	The existing studies on consumer search agree that consumers 
	are worse-off when they do not observe sellers' production 
	marginal cost than when they do.  In this paper we challenge 
	this conclusion.  Employing a canonical model of 
	simultaneous search, we show that it may be favorable for 
	consumer to not observe the production marginal cost.  The 
	reason is that, in expectation, consumer search more 
	intensely when they do not know sellers' production marginal 
	cost than when they know it.  More intense search imposes 
	higher competitive pressure on sellers, which in turn 
	benefits consumers through lower prices.

	
\end{abstract}

\sloppy

\newpage

\section{Introduction}\label{s:intro}

The literature on consumer search has improved our understanding 
of the imperfect competition by assigning a central role to 
consumers' search behavior.  Within this field, one conclusion 
seems well-established: consumers are worse-off when they do not 
observe sellers' production marginal costs than when they 
observe those costs.%
\footnote{The only exception is \cite{janssenetal2017} who show 
\textit{numerically} that the opposite may be true.}
This conclusion holds in a wide range of markets.  Examples 
include markets where exogenous parameters, or shocks, determine 
sellers' production marginal costs (e.g., \cite{dana1994}, 
\cite{janssenetal2011}, \cite{duffieetal2017}), markets where 
consumers make repeated purchases (e.g., \cite{yangye2008}, 
\cite{tappata2009}), and  \textit{vertical industries} where 
sellers buy production inputs from profit maximizing 
manufacturers and, therefore, manufacturers' wholesale prices 
determine sellers' production costs (e.g., 
\cite{janssenshelegia2015}, \cite{garciaetal2017}, 
\cite{janssen2020}).  Moreover, the 
main result of these studies seems to be independent of the 
search protocol.  For instance, \cite{dana1994} and  
\cite{janssenshelegia2015} employ models of sequential search, 
whereas \cite{yangye2008} and \cite{tappata2009} apply models of 
simultaneous search. Also the empirical papers seem to 
apply this conclusion to explain their results (e.g., 
\cite{tappata2009}).

This paper challenges the widespread consensus that consumers 
benefit from observing sellers' production costs. To demonstrate 
that, we employ a canonical search model of 
\cite{burdettjudd1983}.  It is a one shot-game where oligopoly 
sellers simultaneously set prices and consumers choose their 
search intensity.  Sellers are identical, produce homogeneous 
goods and compete on prices.  The production marginal cost in 
the industry is a random draw from a commonly known 
distribution.  Sellers observe the production cost but consumers 
do not, which results in the information asymmetry.  Without 
knowing prices (in addition to the production marginal cost), 
consumers need to engage in costly nonsequential search, also 
known as \textit{simultaneous search} and 
\textit{fixed-sample-size} search, and discover a price of at 
least one firm in order to make purchases.  Each consumer 
chooses a number of firms to search, searches exactly the chosen 
number of firms and then terminates the search. We compare 
market outcomes of this model with that 
of a model where consumers observe the production marginal cost.

Our main finding is that consumers are better-off when they do 
not observe the production marginal cost than when they do.  The 
intuition lies in consumers' search behavior.  In expectation 
consumers search more intensely when there is information 
asymmetry on sellers' production marginal costs than when there 
is not.  More intense search means that consumers are more able 
to compare different offers, which in turn implies stronger 
competition.  Strong competition clearly benefits consumers.  
This benefit is so high that their welfare is higher than 
without the information asymmetry although they spend more 
resources on search.

To understand why consumer search more intensely with the
information asymmetry than without it, it helps to consider a 
case without the information asymmetry.  Specifically, buyers 
search intensity is decreasing and concave in the production 
marginal cost.  The concavity is due to the fact that an 
increase in the production marginal cost has two negative 
effects on the level of price dispersion.%
\footnote{For sufficiently small search costs, in equilibrium
	sellers set prices from a symmetric price distribution that 
	is monotone in its compact support with the monopoly price 
	being the highest price for any realization of the 
	production marginal cost.}
The direct effect arises because of the fact that, consumers' 
search behavior being fixed, the gap between the monopoly price 
and the production marginal cost declines with the production 
marginal cost.  This reduces a range of prices that sellers can 
set and, in particular, the support of the equilibrium price 
distribution shortens with the lowest price in the support 
getting closer to the monopoly price. There is also an indirect 
effect: firms take into account consumers' search strategy when 
pricing.  Firms expect the share of price-comparing 
consumers to drop owing to the above direct effect.  The reason 
is that consumers have less incentive to search and compare 
prices if sellers' prices do not differ much.  Firms thus have 
incentive to charge prices closer to the monopoly price, which 
even further reduces price dispersion.  The direct- and 
indirect-effects go hand in hand with reduction of the level of 
price dispersion, which in turn causes concavity of the search 
intensity in the production marginal cost. Owing to this 
concavity, it then follows by Jensen's inequality that the 
expected search intensity when consumers observe the production 
marginal cost is lower than the search intensity for a given 
expected production marginal cost, i.e., the search intensity 
with the information asymmetry.

This result has important implications in real-world markets.  
A set of such markets are over-the-counter markets.  In these 
markets \textit{benchmarks}, such as LIBOR and EURIBOR, 
aggregate and report interbank borrowing rates on the daily 
basis.  They thus serve as a proxy of sellers' production 
costs.  It is argued that benchmarks resolve the information 
asymmetry on the cost of trading between dealers (who are 
sellers in our model) and traders (who are buyers in 
our model) and improve market traders' well-being 
(\cite{duffieetal2017}). However, our main result warns that the 
benchmarks may improve dealers' profit at the expense of 
traders' well-being.  This seems plausible as in many 
over-the-counter markets, benchmarks are created by dealers' 
initiative rather than that of a public institution.

Other examples of real-world markets, where our main result may 
be relevant, include mortgage markets and consumer financial 
markets.  In these markets, as in over-the-counter 
markets, buyers need to spend time to learn sellers' offers.  
Furthermore buyers wish to quickly gather information about 
offers and it takes time to observe the search outcome, which 
makes simultaneous search a more attractive protocol than 
sequential search (\cite{morganmanning1985}).

The rest of the paper is organized as follows.  The next section 
discusses studies that are closely related to our paper.  In 
Section \ref{s:model} we lay out the model.  In sections 
\ref{s:equil} and \ref{s:cs} we undertake equilibrium analysis 
and compare market outcomes with the information asymmetry on 
the production marginal cost to those without it, respectively.  
In Section \ref{s:extensions} we show that our main results are 
robust to different model extension.  The final section 
concludes.

\section{Related Literature}\label{s:literature}

Our paper contributes mainly to literature on consumer search 
with information asymmetry on the production marginal cost.  
Within this field, \cite{dana1994} is closest to our study.  The 
author analyzes a model that is identical to ours in all 
aspects, except for the search protocol.  He employs sequential 
search.  Thus at each decision node, a consumer needs to decide 
whether to make a purchase at the lowest observed price (if 
there are any), to search one more firm (if there are any 
non-searched firms left), or drop out of the market.  If 
consumers do not observe sellers' production marginal costs, 
they update their beliefs about the production cost each time 
after discovering a new price according to Bayes' rule.  Sellers 
can then signal their production costs through their prices.  In 
equilibrium sellers partially trick consumers into believing 
that their production marginal cost is high by charging high 
prices even though the production marginal cost is 
low.  This trick does not work if consumers observe the marginal 
cost of production.  As a result, sellers have more market power 
with the information asymmetry on the production marginal cost 
than without it. Whereas \cite{dana1994} considers bimodal 
distribution of production marginal costs, 
\cite{janssenetal2011} extend it to a continuous distribution of 
production marginal costs with a compact support. 
\cite{duffieetal2017} further extend the model by making the 
first search costly. Finally, \cite{janssenetal2017} consider a 
duopoly version of \cite{dana1994}'s model and employ a 
so-called \textit{non-reservation price equilibrium} which 
resolves a problem of equilibrium non-existence for certain 
parameter values.  The authors report that in non-reservation 
price equilibrium, the information asymmetry can be beneficial 
for consumers, but this result is based on numerical 
simulations. 

Sellers in our model do not have opportunity to signal their 
production costs via prices because the search is simultaneous.  
The reason is that in a model of simultaneous search, search 
intensity is independent of the prices discovered along the 
search path.

Another set of related studies include papers which examine 
dynamic models of search where consumers make repeated 
purchases.  \cite{tappata2009} employs a model identical 
to ours, but the static game is played over infinitely many 
periods.  The industry production marginal cost evolves 
according to Markov process across periods.  As production costs 
are correlated across periods and consumers observe past 
realizations of production costs, they form expectations on the 
current realization of the production cost. If they expect the 
current production marginal cost to be high, they search less 
intensely, and if they expect low marginal cost of production 
they search more intensely.  The reason is that the level of 
price dispersion is lower with a high production marginal cost 
than with a low cost, as the range of prices that sellers can 
set---measured by the gap between the production marginal 
cost and the monopoly price---is lower in the former case than 
in the latter case.  As a result of this, prices increase fast 
when the production marginal cost rises but decrease slowly when 
the cost falls---a phenomenon known as 
\textit{rockets-and-feathers}.  In contrast if consumers 
observed current realization of the production marginal cost, 
prices would have adjusted instantly to both an increase and a 
decrease in the cost. Therefore, consumers are worse-off when 
they do not observe the production marginal cost than when they 
do.  \cite{yangye2008} study a variation of 
this model where consumers do not observe even the past 
realizations of the production marginal cost.%
\footnote{\cite{cabralfishman2012} propose a dynamic search 
	model with groups types of firms to explain the 	
	rockets-and-feather phenomenon.  However, this result occurs 
	only when changes in production costs over time are small.  
	If such changes are large, the authors show, an opposite 
	results holds: prices adjust to a decrease in the production 
	marginal cost and do not respond to an increase in the 
	production marginal cost.  Hence, the overall impact of the 
	information asymmetry on consumer welfare is ambiguous.}

In our one-shot game, a mechanism discussed in the previous 
paragraph is clearly absent.  It is, however, possible to extend 
our model to a dynamic setting where sellers' marginal costs of 
production are not correlated across periods. In this case our 
main result holds.

Our paper is also related to papers on consumer search in 
vertical industries with downward sloping consumer demand.  
\cite{janssenshelegia2015} introduce a monopolist manufacturer 
to a sequential search model as in \cite{dana1994}.  The 
manufacturer's wholesale price is sellers' production marginal 
costs as sellers buy the manufacturer's product and convert it 
into a final good. Consumers do not observe the wholesale price 
of the manufacturer, i.e., they do not know sellers' production 
marginal costs.  However, in equilibrium consumers have correct 
beliefs about the wholesale price.  The traditional notion is 
that the problem of double marginalization should disappear if 
there are multiple sellers in the market.  However, 
\cite{janssenshelegia2015} show that if consumers have 
arbitrarily small search costs (i.e., when they vanish in the 
limit), manufacturer chooses a wholesale price that is higher 
than the standard monopoly price.  The reason is that if a 
manufacturer deviates to a wholesale price higher than one that 
is conjectured by consumers, sellers cannot increase their 
prices without inducing more search.  If sellers 
were to raise their prices as a response to a higher wholesale 
price, consumers would believe that it must be sellers 
that deviated (and not the manufacturer).  Thus, consumers would 
search further after observing a high price.  This inability of 
sellers to raise prices as the wholesale price increases makes 
manufacturers' demand less responsive, which allows the 
manufacturer to squeeze out sellers' profits.  As a result, the 
equilibrium wholesale price is higher than the standard monopoly 
price.  This clearly harms consumers. Note however that such a 
mechanism would certainly be absent if consumers observed the 
wholesale price. \cite{garciaetal2017} extend the model by 
assuming that there are multiple manufactures and sellers, just 
like consumers, need to engage in costly search in order to 
discover wholesale prices. \cite{janssen2020} allows a 
manufacturer to employ two-part tariffs and 
\cite{janssenreshidi2020} permit a manufacturer to price 
discriminate among sellers.%
\footnote{A polemical assumption in these studies is
	that, although consumers do not know sellers' production 
	marginal costs, they know manufacturers' production costs.}

Finally, \cite{sobolev2017} studies search markets where each 
firm's marginal cost of production is an independent random 
draw.  As each firm observes only its own production marginal 
cost, there is no such information asymmetry between the sellers 
and buyers as we have it in our paper. The author shows that 
firms can raise their profits by submitting their production 
costs to a benchmark which publishes the 
average of the submitted costs.  The author shows that the 
driving force of this result is partial resolution of production 
cost uncertainty among sellers.  Unlike \cite{sobolev2017} we 
focus on the role of benchmarks that resolve information 
asymmetry between buyers and sellers.

\section{Model}\label{s:model}

Our model is an extension of \cite{burdettjudd1983}. $N\geq 2$ 
number of identical firms supply homogeneous goods to a unit 
mass of consumers.  Production marginal cost is a random draw 
from $\left\{c_1, c_2, ... , c_K\right\}$, where $0 \leq c_1 
\leq ... \leq c_K < \infty$, according to probability mass 
function (PMF) $f$, so that $f_k$ represents the probability 
that the production marginal cost equals to $c_k$ for $k \in 
\{1,...,K\}$.  Sellers observe their production costs and 
compete on prices.  Price $p^j(c_k)$ represents firm $j$'s 
strategy for given production cost $c_k$.  As we allow for mixed 
strategies, let $x^j_k(p)$ be the probability that firm $j$ 
charges a price above $p$ and the support of the price 
distribution be $[\underline{p}_k, \overline{p}_k]$, when the 
production marginal cost is $c_k$.

Each consumer wishes to consume a unit of a product which she 
values at $v(>c_K)$.  Buyers, unlike sellers, do not observe the 
production marginal cost.  They also do not know prices.  To 
make a purchase a consumer has to learn at least one price 
through costly search.  Search is simultaneous.  A consumer 
chooses $m$ number of firms to search and having searched all 
$m$ firms she terminates the search.  It costs $s$ to obtain a 
price quote. To ensure full participation we let searching one 
firm to be free.  Therefore, searching $m$ number of firms 
entails a total cost of $(m-1)s$.  Since mixed-strategies are 
allowed, we let $q(m)$ stand for the probability that a buyer 
searches $m$ firms and so $\sum_{m=1}^{N}q(m)=1$.%
\footnote{As searching one firm is free, not searching at all is
	always a weakly dominated strategy.}

The timing of the game is as follows.  First, the production 
marginal cost is realized. Second, firms observe the production 
marginal cost and simultaneously set prices.  Third, 
consumers---without knowing prices and production marginal 
cost---search.  Consumers who observe at least one price 
may make purchases.  

The solution concept is Bayesian-Nash equilibrium (BNE).  
Letting $x^{-j}_k$ be the strategies of all firms other than 
firm $j$, $\Pi^j_k(p,x^{-j}_k)$ be the expected profit that 
price $p$ yields to firm $j$ and $\overline{\Pi}_k \geq 0$ be 
some constant, a BNE is a collection of price distributions 
$(x^j_k)_{j=1}^N$ for each $c_k$ and search probabilities 
$q(m)$ such that (i) $\Pi^j_k(p,x^{-j}_k) \leq \overline{\Pi}_k$ 
for all $p$, (ii)  $\Pi^j_k(p,x^{-j}_k) \geq \overline{\Pi}_k$ 
for $p \in [\underline{p}_k, \overline{p}_k]$, and (iii) 
consumers searching for $m$ firms do not obtain lower utility 
than searching for any number of firms.

\section{Equilibrium}\label{s:equil}

Equilibrium analysis consists of two parts. In the first part we 
examine our main model where buyers do not observe the 
production marginal cost.  We later analyze a model which 
differs from our main model in that buyers observe the 
production marginal cost.

\subsection{Unobserved Production Marginal Cost}

We start the equilibrium analysis by noting that there always 
exists an equilibrium where all buyers search one firm and all 
firms charge the monopoly price (\cite{diamond1971}), a result 
is known as the \textit{Diamond paradox}.  The reasoning is as 
follows. Suppose that buyers expect all firms to price their 
products at the monopoly price.  It is then optimal for buyers 
to search one firm.  If, however, buyers do not compare prices, 
the monopoly price is the optimal price, which justifies buyers' 
above expectation.

\begin{proposition}\label{prop:diamond}
	There always exists a BNE where buyers search one firm and 
	firms charge the monopoly price. 
\end{proposition}

The Diamond paradox is known to be fragile to model assumption.  
It ceases to exist if, for example, one introduces a small share 
of price-comparing buyers.  We thus turn our attention to BNEs 
where some buyers search more than one firms, which we call a 
BNE with active search.  Existence of such equilibrium is robust 
to small changes in model assumption.  

Our first observation is that in equilibrium with active search 
a positive share of consumers search only one firm.  The 
reasoning is by contradiction. If all buyers observe at least 
two prices, they will buy at the lowest observed price.  No firm 
then wishes to be one with the highest price. If two or more 
firms tie at some price higher than the production marginal 
cost, it pays off for one of them to slightly undercut its 
price.  As a result of these two arguments, the optimal price 
must be equal to the production marginal cost.  However, if 
prices equal to the production marginal cost, buyers do not want 
to search more than one firm, a contradiction. 

\begin{lemma}\label{lem:q1}
	In a BNE with active search, some buyers must search one 
	firm.
\end{lemma}

A consequence of the lemma is that in equilibrium with active 
search, firms play mixed-strategy pricing for any realization of 
the production marginal cost (e.g., \cite{bayeetal1992}). This 
happens because of two opposing forces that affect a firm's 
pricing policy.  One of them is due to price-comparing buyers 
whose share must be strictly positive in equilibrium with active 
search.  The opposing force is due to buyers who search only one 
firm and whose share is also positive as Lemma \ref{lem:q1} 
shows. Each firm wishes to raise its price to ripoff buyers who 
observe only its offer, but at the same time each firm wishes to 
lower its price to attract price-comparing consumers.  An 
interaction of these two opposing forces gives rise to a 
mixed-strategy pricing.

If buyers expect dispersed prices and searching one firm is part 
of an equilibrium with active search, price-comparing consumers 
search exactly two firms in equilibrium. To understand the 
reasoning one needs to compare the added benefit of searching 
an additional firm to its cost. The added benefit of 
searching an additional firm decreases with the number of 
searched firms for any non-degenerate price distribution.  
Formally, if we let $X(p)$ be the \textit{ex-ante} probability 
that consumers expects a firm to charge a price above $p$, then
$1-(1-X(p))^m$ is the distribution of the minimum of $m$ prices. 
Call is $m$th order statistic.  The difference between the $m$th 
and $m+1$th order statistic is then $X^m(p)(1-X(p))$.  This 
difference is clearly decreasing in $m$, which proves that the 
added benefit of searching an additional firm is decreasing with 
the number of searched firms. However, as the cost of searching 
an additional firm is constant, it must be that either all 
consumers search the same number of firms or they must randomize 
over searching two adjacent numbers of firms.  Since in 
equilibrium (with active search) some consumers discover only 
one price and the rest of them compare different prices, these 
price-comparing consumers must search exactly two firms. 

\begin{lemma}\label{lem:q1q2}
	In a BNE with active search, some of consumers searche one 
	firm and the rest, two firms.
\end{lemma}

The lemma helps us to determine properties of the equilibrium 
price distribution for each realization of the production 
marginal cost. First, it has been established that for any given 
production marginal cost, firms draw prices from the same unique 
price distribution (see, e.g., \cite{burdettjudd1983}, 
\cite{johnenronayne2020}).%
\footnote{The full proof is given by \cite{johnenronayne2020}. To
	understand the intuition, consider a case where
	price comparing consumers observe three prices instead to 
	two prices.  This means that each firm competes with two 
	other firms for the price-comparing buyers.  It is then 
	possible to show that it is optimal for one of the firms to 
	always charge the monopoly price, while the other two 
	compete head-to-head by drawing prices from the same 
	distribution that continuously increases in its support. 
	However, if price comparing buyers observe exactly two 
	prices, then each firm has to compete head-to-head to every 
	other firm. This eliminates a firm's incentive to always 
	charge the monopoly price, as its rivals will always 
	undercut.}
We will thus drop firm-specific indices from the equilibrium 
price distributions.  Second, the price distribution cannot not 
have atoms.  If it did have an atom, undercutting would be 
profitable.   Third, the price distribution cannot have flat 
regions in the support.  If it did, an individual firm would 
strictly prefer the highest price in that flat region to the 
lowest price in the same region as its expected demand would be 
the same at those prices. Fourth, the highest price in the 
support of the equilibrium price distribution must be equal to 
the monopoly price $v$.  If the highest price was greater than 
the monopoly price, a firm does not make any sales at that 
price.  If the highest price was lower than the monopoly price, 
a firm can profitable deviate to the monopoly price as in both 
cases it sells only to buyers who observe its price only.

In the light of those four properties we are now ready to derive 
the equilibrium price distribution for a given production 
marginal cost.  For a given $c_k$, if firm $j$ charges price 
$p$, its expected profit is
\begin{equation*}
	\Pi_j(p,x_{k}^{-j}) = \left(\frac{1-q^U}{N} + \frac{2 
	q^U}{N} x_k(p)\right) (p-c_k),
\end{equation*}
where we let $q^U\equiv q(2)$ and so $q(1) = 1-q^U$ to simplify 
the notation.  The first term in the large brackets represent 
the share of consumers that search one firm and happen to visit 
the firm under question.  These consumers make a purchase at any 
price below the monopoly price.  The second term in the large 
brackets stands for the share of consumers who search two firms 
and happen to visit the firm under question as well as another 
competitor.  These consumers buy firm $j$'s product if its price 
is lower than the rival firm's price (as well as the monopoly 
price).  

As any firm is indifferent of choosing any price 
in the support of the equilibrium price distribution and prefers 
these price to those which are not in the support, we equalize 
the above expected profit to the expected profit that the 
monopoly price and obtain
\begin{equation}\label{eq:xk}
	x_k(p) = \frac{1-q^U}{2q^U} 
	\left(\frac{v-c_k}{p-c_k}-1\right) \ 
	\mbox{with support} \ 
	\left[\underline{p}_k, v\right],
\end{equation}
where $\underline{p}_k$ solves $x_k(\underline{p}_k)=1.$ It is 
useful to work with inverse function $p_k(x)$ which in 
equilibrium satisfies $p_k(x) = (1-q^U)(v-c_k)/(1-q^U + 
2q^Ux)+c_k$.

It is now left to check whether an individual consumer is 
indifferent between searching one firm and searching two firms 
given the price distributions in \eqref{eq:xk}.  Searching one 
firm yields an expected payoff equal to 
\begin{equation*}
	v + \sum_{k=1}^{K}f_k \int_{\underline{p}_k}^{v}px_k'(p)dp = 
	v - \sum_{k=1}^{K}f_k \int_{0}^{1}p_k(x)dx,
\end{equation*}
where we changed the variable of integration to obtain the 
equality.  Searching two firms yields an expected payoff equal to
\begin{equation*}
	v + 2 \sum_{k=1}^{K}f_k \int_{\underline{p}_k}^{v} p  x_k(p) 
	x_k'(p)dp - s = v - 2 \sum_{k=1}^{K}f_k \int_{0}^{1} p_k(x)  
	x dx - s.
\end{equation*}
These two payoffs must be equal for an individual buyer to be 
indifferent between searching one firm and searching two firms.  
Equalizing the payoffs we obtain
\begin{equation}\label{eq:qU}
	\sum_{k=1}^{K}f_k \int_{0}^{1} p_k(x) 
	(1-2x)dx =s.
\end{equation}
The challenge is then to show that there exists $0<q^U<1$ that 
solves this equation. The following proposition demonstrates 
that such $q^U$ exists for small search costs.

\begin{proposition}\label{prop:eq_asym}
	Suppose buyers do not observe the production marginal cost. 
	Then, for any $N\geq 2$ and $f$, there exists 
	$\overline{s}>0$ such that for $s< \overline{s}$ there 
	exist two BNEs with active search given by 
	$((x_k)_{k=1}^{K},q^U)$ where $x_k$ 
	is determined by \eqref{eq:xk} for each $k\in \{1,...,K\}$ 
	and $q^U$ by \eqref{eq:qU}.
\end{proposition}

\begin{figure}[ht]
	\centering
	\includegraphics[width=.7\linewidth]{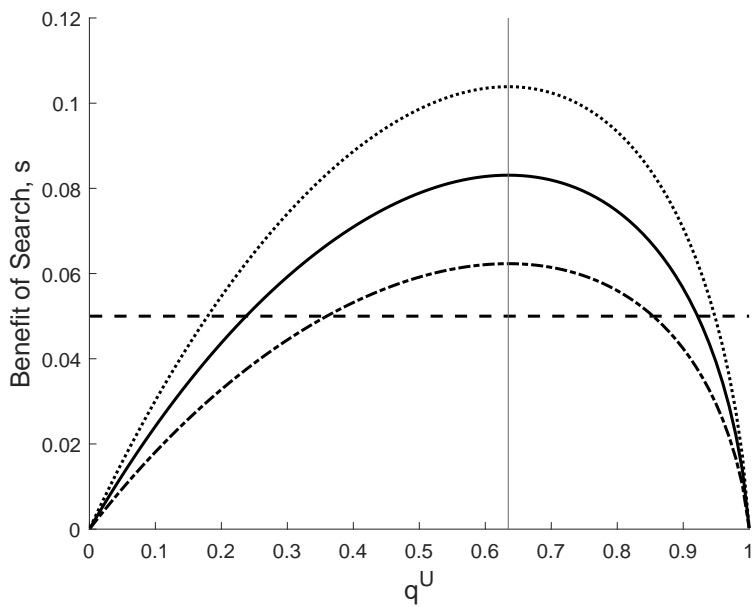}  
	\captionsetup{justification=centering}  
	\caption{Illustration of BNEs for $N=2$, $v=1$, $s=0.05$, 
		$K=2$, $c_1=0$, $c_2=0.4$ and $f_1=0.5$.}
	\label{fig:eq_u}
\end{figure}

The proof is in the appendix and the intuition is as follows.  
In the appendix we show that the expected benefit of searching 
the second firm, which is given by the left-hand side of 
\eqref{eq:qU}, is positive and concave in the search intensity 
$q^U \in (0,1)$.  Moreover, the expected benefit of searching 
the second firm vanishes as the share of price-comparing 
consumers either disappears or converges to one.  Intuitively, 
this makes sense as firms have incentive to charge the monopoly 
price if the share of price-comparing buyers vanishes (recall 
the Diamond paradox) and the price equal to the production 
marginal cost if all buyers compare prices (recall the argument 
behind Lemma \ref{lem:q1}).  These facts imply that the expected 
benefit of searching the second firm is inverse U-shaped with 
respect to the search intensity as illustrated by the solid 
curve in Figure \ref{fig:eq_u}. Then, for small search costs 
there exist two equilibria.   In Figure \ref{fig:eq_u} these 
equilibria are represented by the intersections of the solid 
curve and the dashed line representing the search cost.  

One can argue that only one of the two equilibria is stable in a 
sense that if the actual search intensity is in the neighborhood 
of the equilibrium search intensity, then consumers optimally 
adjust their search intensity to the equilibrium one.%
\footnote{Formally, let $A_{\varepsilon} =
	\left\{\widetilde{q} \in \mathbb{R}: |\widetilde{q} - 
	q^U|<\varepsilon\right\}$ be the neighborhood of an 
	equilibrium $q^U$ for $\varepsilon>0$ and $\varepsilon \to 
	0$.   Then a perturbed search intensity $q'$ is any search 
	intensity in that neighborhood, i.e., $q' \in 
	A_{\varepsilon}$.  If $q^U$ is a part of a stable 
	equilibrium and the actual search intensity is $q'$, the 
	search intensity converges to the equilibrium one.}
The question is then, which equilibrium is stable?  The 
following corollary provides an answer.

\begin{corollary}
	Of the two BNEs in Proposition \ref{prop:eq_asym}, one
	characterized by a higher search intensity is stable.
\end{corollary}

The reasoning is as follows. Consider an equilibrium given by a 
high search intensity, i.e., the right-most intersection of the 
two curves in Figure \ref{fig:eq_u}. If the actual search 
intensity falls slightly short of the equilibrium one, the 
expected benefit of searching the second firm is higher than 
the cost of doing so.  Therefore, buyers have incentive to 
search more intensely.  If, in contrast, the actual search 
intensity is higher than the equilibrium one, the expected 
benefit of searching the second firm is lower than the cost of 
doing so.  As a result, buyers have incentive to search less 
intensely.  By applying similar arguments, it is easily 
established that the equilibrium characterized by a lower search 
intensity is unstable.

\subsection{Observed Production Marginal Cost}

We now turn to examining a case where both firms and buyers 
observe the production marginal cost.  This allows buyer to 
condition their search strategies on the realized production 
marginal cost. If the marginal cost of production is $c_k$, the 
resulting \textit{ex-interim} game is a special case of our main 
model where buyers do not observe the production marginal cost 
and $f_k=1$.  Therefore, we employ Nash equilibrium (NE) as a 
solution concept.  Also we can apply our analysis in the 
previous subsection to analyze the case where buyers observe the 
realization of the production marginal cost.  To avoid 
repetition, we omit parts of the analysis which directly follow 
from those in the previous subsection.

Let $q(c_k)$ represent the probability that consumers search two 
firms when the marginal cost of production is $c_k$, and so the 
probability that buyers search one firm is $1-q(c_k)$.  Then, 
the equilibrium consists of $(x_k,q(c_k))$ for each realization 
of $k \in \{1,...,K\}$.  Following the line of argument in the 
previous subsection, we can establish that
\begin{equation}\label{eq:xk_sym}
	x_k(p) = \frac{1-q(c_k)}{2q(c_k)} 
	\left(\frac{v-c_k}{p-c_k}-1\right) \ 
	\mbox{with support} \ \left[\underline{p}_k, 
	\overline{p}_k\right],
\end{equation}
where $\underline{p}_k$ solves $x_k(\underline{p}_k)=1$.  Using 
an inverse function $p_k(x)$ that satisfies $p_k(x) = 
(1-q(c_k))(v-c_k)/(1-q(c_k) + 2q(c_k) x)+c_k$ in equilibrium, we 
write an equation that determines the equilibrium search 
intensity:
\begin{equation}\label{eq:qk}
	\int_{0}^{1} p_k(x) (1-2x)dx =s.
\end{equation}
We are now ready to state the main result of this subsection in 
the following corollary, which is a direct consequence of 
propositions \ref{prop:diamond} and \ref{prop:eq_asym}.

\begin{corollary}\label{cor:eq_sym}
	Suppose buyers observe the production marginal cost. 
	\begin{itemize}
		\item[(i)]  There always exists the Diamond-paradox 
		equilibrium.
		\item[(ii)] For each $k = \{1,...,K\}$, there exists 
		$\overline{s}_k>0$ such that for $s\leq \overline{s}_k$ 
		there exist two NEs with active search given by 
		$(x_k,q(c_k))$ which are determined by \eqref{eq:xk_sym} 
		and \eqref{eq:qk}.
		\item[(iii)] Of the two NEs with active search for each 
		$k \in \{1,...,K\}$, one with a higher search intensity 
		is stable.
\end{itemize}
\end{corollary}

Two questions arise naturally.  One is, how $\overline{s}_k$s 
are related to each other? The other is, how $\overline{s}_k$s 
are related to $\overline{s}$ in Proposition 
\ref{prop:eq_asym}?  The following corollary answers the both 
questions.

\begin{corollary}\label{cor:sk}
	We have (i) $\overline{s}_1 > \overline{s}_2> ... > 
	\overline{s}_K$ and (ii) $\overline{s} = \sum_{k=1}^K f_k 
	\overline{s}_k$.
\end{corollary}

The proof is in the appendix and the intuition is as follows. 
Recall from the discussion of Proposition \ref{prop:eq_asym} 
that the added (expected) benefit of searching the second firm 
is concave in the search intensity (also observe Figure 
\ref{fig:eq_u}).  This means that there is a unique value of the 
search intensity which maximizes the added benefit of searching 
the second firm.  In the appendix we show that this unique value 
of the search intensity, which maximizes the added benefit of 
searching the second firm, is independent of the production 
marginal cost.  Moreover, the maximum value of the added benefit 
of searching the second firm is linearly decreasing in the 
marginal cost of production.  This last observation implies the 
first part of the corollary and, along the observation before 
it, also results in the corollary' second part.

Figure \ref{fig:eq_u} helps to illustrate this point.  The 
dotted and dash-dotted curves represent the added benefit of 
searching the second firm when the production marginal cost is 
low (i.e.,$c_1=0$) and when it is high ($c_2=0.1$), 
respectively, and when consumers observe them.  The vertical 
line stands for the search intensity at which the added benefit 
of searching the second firm is maximized in these two cases as 
well as in the case with the information asymmetry.

\section{Welfare Analysis}\label{s:cs}

We are now in a position to examine the role of the information 
asymmetry on the production marginal cost: firms observing their 
marginal cost of production and buyers not observing it.  The 
following proposition states the main result of this section.

\begin{proposition}\label{prop:welfare}
	Suppose $s \leq \overline{s}_{K}$ and consider stable 
	equilibria with active search.  Then, in expectation
	\begin{itemize}
		\item[(i)] consumers search more,
		\item[(ii)] the consumer surplus is higher,
		\item[(iii)] the firm profit is lower, and
		\item[(iv)] the total surplus is lower,
	\end{itemize}
	when consumers do not observe the marginal cost of 
	production than when they do.
\end{proposition}

The proof is in the appendix.  To understand the intuition 
behind (i), it is useful to consider consumers' search intensity 
when they observe the production marginal cost.  In the appendix 
we show that this search intensity is decreasing and concave in 
the production marginal cost.  The reasoning is as follows.  As 
the production marginal cost rises while all else remains the 
same, the level of price dispersion shrinks.  This is because 
the range of prices firms can set---which is the difference 
between the monopoly price and the production marginal 
cost---decreases and, in particular, it gets closer to the 
monopoly price.  A low level of price dispersion means that 
prices do not differ much across sellers.  This in turn implies 
that consumers have less incentive to search and compare 
prices.  If the share of price-comparing consumers falls, 
sellers have incentive to charge high prices, namely prices 
closer to the monopoly price.  As a result, price dispersion 
shrinks even more, which in turn causes less search. This 
explains why consumers' search intensity falls fast as the 
production marginal cost rises, i.e., why it is decreasing and 
concave with respect to the production marginal cost.  

By Jensen's inequality it then follows that the expected search 
intensity when consumers observe the marginal cost of production 
is lower than the search intensity given the expected production 
marginal cost.  Formally we have that $\mathbb{E}[ q(c_k)] < 
q\left(\mathbb{E}[c_k]\right)$ where the expectation is with 
respect to the production marginal cost. However, we know that 
in our model with information asymmetry, consumers take into 
account the expected production marginal cost when deciding on 
their search intensity.  Then, the search intensity under the 
information asymmetry equals to $q\left(\mathbb{E}[c_k]\right)$, 
meaning that consumers search more intensely with the 
information asymmetry than without it. 

Intuition behind (ii) and (iii) is easily understood together. 
Recall that the equilibrium firm profit equals to $\sum_{k=1}^K 
f_k (1-q(c_k))(v-c_k)/N$ when consumers observe the production 
marginal cost and to $\sum_{k=1}^K f_k (1-q^U)(v-c_k)/N$ when 
they do not.  Also firms' market power is measured by 
$(1-q(c_k))/(2 q(c_k))$ and $(1-q^U)/(2 q^U)$ respectively in 
models without and with the information asymmetry.  Notice that 
the equilibrium profits and firms' market power decrease with 
the respective search intensities.  This makes sense since 
competition gets stronger as the share of price-comparing 
consumers rises.  From (i) we know that $q^U = 
q\left(\mathbb{E}[c_k]\right)$ and thus the expected search 
intensity with the information asymmetry is higher than that 
without it.  Since competition is more intense with higher 
search intensity, firms' market power is (in expectation) lower 
with the information asymmetry than without it.  Stronger 
competition is detrimental for firm profit but clearly benefits 
consumers as they make purchases at lower prices.  

The reasoning behind (iv) follows directly from (i).  Notice 
that if all consumers make purchase, the total surplus depends 
only on the total costs spent on search.  From (i) we know that 
consumers search more intensely and hence incur higher search 
costs in total when they do not observe the production marginal 
cost than when they do.  Therefore the total surplus is lower in 
the former case than in the latter.

Proposition \ref{prop:welfare} has an important implication on 
firms' incentive to disclose information on the production 
marginal cost.  It implies that firms have incentive to disclose 
the realization of the production marginal cost.  

\begin{corollary}\label{cor:disclosure}
	Suppose that $s \leq \overline{s}_{K}$, consider stable 
	equilibria with active search, and assume that information 
	disclosure is free.  If consumers do not observe 
	the production marginal cost, firm have incentive to 
	disclose such information to consumers. 
\end{corollary}

The reasoning is as follows.  It is clear that firms choose to 
disclose their production marginal cost if they need to 
simultaneously decide whether to do so \textit{before} the 
realization of the production marginal cost.  If, however, firms 
can choose their disclosure strategy \textit{after} the 
realization of the production marginal cost, we can employ the 
following algorithm to show ``full unraveling.'' Consider first 
a case where the production marginal cost obtains its higher 
value.  If this information is not disclosed to consumers, their 
search intensity is given by $q^U$ as in Proposition 
\ref{prop:eq_asym}.  If, in contrast, firms inform buyers about 
the production marginal cost, we know that buyers search less 
intensely than $q^U$.  Since firms market power is higher when 
they disclose the information than when they do not, they choose 
to disclose.  Consider next a case where the production marginal 
cost obtains its second highest value.  Just like in the 
previous case, firms have incentive to disclose this information 
to mitigate search.  We can continue the argument in a similar 
manner to see that firms have incentive to disclose information 
for all, but the lowest, value-realizations of the production 
marginal cost.  Consumers correctly conjecture that if there is 
no information disclosure, the production marginal cost must 
have obtained its lowest value.

We next provide sufficient conditions under which the resolution 
of the information asymmetry between sellers and consumers 
improves consumers' well-being.  

\begin{proposition}\label{prop:welfare_reverse}
	Suppose $s > \overline{s}$ meaning that there is no 
	equilibrium with active search when consumers do not observe 
	the production marginal cost.  Then, in expectation
	\begin{itemize}
		\item[(i)] consumers search (weakly) less,
		\item[(ii)] the consumer surplus is (weakly) lower,
		\item[(iii)] the firm profit is (weakly) higher, and
		\item[(iv)] the total surplus is (weakly) higher
	\end{itemize}
	when consumers do not observe the production marginal cost 
	than when they do.  All these results hold strictly if 
	$s<\overline{s}_1$.
\end{proposition}

The reasoning is fairly simple.  If the search cost is high 
enough, i.e.,  $s > \overline{s}$, the only equilibrium with the 
information asymmetry is the Diamond paradox.  In that 
equilibrium consumers search one firm and sellers charge the 
monopoly price. However, if consumers observe the production 
marginal cost, there may exist an equilibrium with active search 
for small values of the production marginal cost.  Formally it 
may be that $s<\overline{s}_1$, $s > \overline{s}$ and $f_1>0$.  
This possibility of the equilibrium with active search results 
in lower profit and higher consumer welfare when consumers 
observe the production marginal cost.

This result is reminiscent of the main result in 
\cite{duffieetal2017} when it comes to the search intensity. In 
line with our result in Proposition \ref{prop:welfare_reverse} 
the authors show that consumers may search more intensely when 
they observe the production marginal cost than when they do not, 
given a moderately high search cost.

\section{Extensions}\label{s:extensions}

With this section we show that our main result---consumers 
benefiting from information asymmetry on the production marginal 
cost---is robust to different model extensions.  We present 
three different variations of the model.  In the next subsection 
we assume that the production marginal cost is random draw from 
atomless distribution with compact support. Subsection 
\ref{ss:truly_costly} presents a case where searching each firm 
is costly, including the first firm. The final subsection 
incorporates into the main model a small share of 
price-comparing consumers. 

We relegate extensive analysis of the final two subsection to 
the appendix in order to avoid repetition.  In our analysis we 
do not establish uniqueness of a stable equilibrium with active 
search, although we focus on such equilibria.

\subsection{Continuous Distribution of Production Marginal 
Cost}\label{ss:continuous}

We next extend our model to continuous distribution of 
production marginal cost.  We assume that $F$ has no mass points 
or gaps in its compact support $[\underline{c}, \overline{c}]$ 
where $0\leq \underline{c}<\overline{c}< v$. The rest of the 
model remains unchanged.

For equilibrium analysis we can employ the same line of argument 
as in Section \ref{s:equil}.  Correspondingly we can also employ 
the same techniques to prove the main results.  The only 
qualitatively inconsequential difference we need to take care of 
is a replacement of summation signs with corresponding integral 
signs whenever we wish to take an expectation with respect to 
the production marginal cost. As a result, all our results in 
the main model follow: a stable equilibrium with active search 
existing for small search costs both when consumers observe the 
production marginal cost and when they do not; and the 
information asymmetry on the production marginal cost benefiting 
consumers.

\subsection{Truly Costly Search}\label{ss:truly_costly}

In the main model we assumed that searching one firm was free in 
order to ensure full consumer participation.  In this subsection 
we presume that searching any firm entails a search cost of $s$.%
\footnote{\cite{janssenetal2005} is the first paper to examine
	the first search being costly in a sequential search model.}
Formally, if a consumer searches $m$ firms she then incurs a 
total cost of $m s$.  Therefore it may happen that some 
consumers decide to not search at all in equilibrium.

This additional change in the model does not affect our main 
results.  Existence of stable equilibria with active search is 
determined by the same conditions as in the main model.  
Our main result in Proposition \ref{prop:welfare}---that 
consumers are better-off with the information asymmetry than 
without it for small search costs---remains true.

There is only one additional condition that is required to 
ensure existence of stable equilibria with active search.  We 
need to show that in stable equilibria with active search, 
consumer prefer searching one or two firms to not searching at 
all.  This is indeed the case as we demonstrate it in Appendix 
\ref{aa:trulycost}.

Each search being costly is qualitatively consequential if the 
search cost is high.  Consider, for example, a case where 
consumers do not observe the production marginal cost and the 
search cost is so high that equilibrium with active search does 
not exist, i.e., $s > \overline{s}$.  In equilibrium  consumers 
do not search at all.  This is different from equilibrium of our 
main model where consumers search one firm and make purchases at 
the monopoly price---the Diamond paradox.  The reason why 
consumers do not search with each search being costly is as 
follows. If consumers do not search and do not compare prices, 
firms optimally set the monopoly price.  If, however, firms 
charge the monopoly price, it is optimal for consumer to not 
search and this justifies firms' expectation that consumers do 
not compare prices. 

Due to this different characteristic of the Diamond paradox, 
some of our results in Proposition \ref{prop:welfare_reverse} 
does not hold in the current extension. Specifically for high 
search costs, i.e., $s > \overline{s}$, there is no trade when 
consumers do not observe the production marginal cost. As a 
consequence the total surplus and firm profit are (weakly) lower 
when consumers do not observe the marginal cost of production 
than when they do. This result is in line with the main result 
in \cite{duffieetal2017}.

\subsection{Price-comparing Consumers}\label{ss:compare}

In this subsection we show that our main results do not 
change qualitatively if we introduce a small share of 
price-comparing consumers. Specifically we assume that a 
positive but small share of consumers, given by $\lambda>0$, 
always compare multiple prices. For simplicity we let these 
consumers observe two prices and buy outright at the 
lowest of the observed prices.%
\footnote{It is generally possible to let price-comparing
	consumers to observe any number of multiple prices (and at 
	most $N$ number of prices). Independent of exactly how many 
	price these consumers compare, the market becomes more 
	competitive as the share of price-comparing buyers 
	increases.  In an extreme case if the share of 
	price-comparing consumers vanishes, the model collapses to 
	our main model and the market outcomes are the same as those 
	of the main model.}

In Appendix \ref{aa:shoppers} we show that the market outcome 
withing the current model converges to those of the main model 
as the share of price-comparing consumers vanishes. It then 
means that for sufficiently small share of price-comparing 
buyers, results of the current model must be qualitatively the 
same as those of the main model.

\section{Conclusion}

We see the paper to be the first to identify the favorable 
impact of information asymmetry on the production marginal cost 
on competition and consumer welfare in search markets. Our 
result does not only challenge a theoretical conclusion that 
seems to be agreed upon in consumer search literature, but 
also has real-world implication markets with small search costs. 

\newpage
\appendix
\numberwithin{equation}{section}

\begin{singlespace}

\section{Proofs}

\subsection{Proof of Proposition \ref{prop:eq_asym}}

To show the existence, we rewrite \eqref{eq:qU} by using 
$p_k(x)$ as
\begin{equation}\label{eq:app_qU}
	 \sum_{k=1}^K f_k \int_{0}^1 
	 \left(\frac{(1-q^U)(v-c_k)}{1-q^U + 2q^U x} + 
	 c_k\right)(1-2x) dx
	= s.
\end{equation}
It then suffices to show the following facts: (i) that the LHS 
of \eqref{eq:app_qU} is positive for any $0<q^U<1$, (ii) is 
strictly concave in $q^U \in (0,1)$, (iii) converges to zero 
both as $q^U\downarrow 0$ and as $q^U \uparrow 1$. The LHS is 
indeed positive as for each $k \in \{1,...,K\}$ we have
\begin{equation*}
	\begin{aligned}
&\int_{0}^1 (1-2x) \left(\frac{(1-q^U)(v-c_k)}{1-q^U + 2q^U 
x} + c_k\right) dx \\
= & \int_{0}^{1/2}	(1-2x) \left(\frac{(1-q^U)(v-c_k)}{1-q^U + 
2q^U x} + c_k\right) dx - \int_{1/2}^1 (2x-1) 
\left(\frac{(1-q^U)(v-c_k)}{1-q^U + 2 q^U x} + c_k\right)dx\\
\geq & \int_{0}^{1/2} (1-2x) \left((1-q^U)(v-c_k) + 
c_k\right) dx - \int_{1/2}^1 (2x-1) \left((1-q^U)(v-c_k) + 
c_k\right)dx\\
= &\left((1-q^U)(v-c_k) + c_k\right) \int_{0}^{1}(1-2x)dx = 0,
\end{aligned}
\end{equation*}
where we set $x=1/2$ in the large brackets to obtain the 
inequality. To establish concavity of the LHS of 
\eqref{eq:app_qU} in $q^U \in (0,1)$, we differentiate 
$(1-q^U)/(1-q^U + 2q^U x)$ twice w.r.t. $q^U$ to obtain	
$-4x(1-2x)/(1 - q^U + 2q^Ux)^3$. Then, the twice derivative of 
the LHS of \eqref{eq:app_qU} w.r.t. $q^U$ is
\begin{equation*}
	- \sum_{k=1}^K f_k \left(v-c_k\right)  \int_{0}^1 \frac{4 x 
	(1-2x)^2 }{(1 - q^U + 2q^Ux)^3}dx,
\end{equation*}
which is strictly negative as the integrand is positive for any 
$0<q^U<1$.  This shows that the LHS of \eqref{eq:app_qU} is 
indeed concave in $q^U \in (0,1)$.  We finally observe that the 
LHS of \eqref{eq:app_qU} converges to zero both as $q^U 
\downarrow 0$
\begin{equation*}
	\sum_{k=1}^K f_k \lim\limits_{q^U \downarrow 0} \int_{0}^1 
	\left(\frac{(1-q^U)(v-c_k)}{1-q^U + 2q^U x} + 
	c_k\right)(1-2x) dx =  \sum_{k=1}^K f_k \int_{0}^1  v(1-2x) 
	dx =0
\end{equation*}
and as $q^U \uparrow 1$
\begin{equation*}
	 \lim\limits_{q^U \uparrow 1}  \sum_{k=1}^K f_k \int_{0}^1 
	\left(\frac{(1-q^U)(v-c_k)}{1-q^U + 2q^U x} + 
	c_k\right)(1-2x) dx =  \sum_{k=1}^K f_k c_k \int_{0}^1 
	(1-2x) dx =0.
\end{equation*}

From facts (i), (ii) and (iii), the proof of the proposition 
immediately follows.

\subsection{Proof of Corollary \ref{cor:sk}}

(i) We start the proof by noting that 
\begin{equation}\label{eq:sk}
	\begin{aligned}
\overline{s}_k &&=&&& \max_{q(c_k)} \left\{\int_{0}^1 
\left(\frac{(1-q(c_k))(v-c_k)}{1-q(c_k) + 2q(c_k) x} + 	
c_k\right)(1-2x) dx\right\}\\
&&=&&& \max_{q(c_k)} \left\{(v-c_k)\left( \ln 
\left(\frac{1+q(c_k)}{1-q(c_k)}\right) - 
2q(c_k)\right)\frac{1-q(c_k)}{2q(c_k)^2}\right\},
	\end{aligned}
\end{equation}
where we simply employed integration to obtain the second line 
and $q(c_k)^2 := [q(c_k)]^2$.  We next establish two facts: (a) 
the unique value of $q(c_k) \in (0,1)$ that maximizes the 
expression in the curly brackets in 
\eqref{eq:sk} does not depend on $c_k$ and (b) $\overline{s}_k$ 
decreases in $c_k$.  To prove (a), we note that the expression 
in the curly brackets in \eqref{eq:sk} is concave in $q(c_k)$ 
which follows from the proof of Proposition \ref{prop:eq_asym}. 
Then, there exists a unique value of $q(c_k) \in (0,1)$, denoted 
by $q^*$, that maximizes that expression in the curly brackets.  
This $q^*$ solves 
\begin{equation*}
	\left.\frac{\partial(v-c_k)\left(\ln 
	\left(\frac{1+q(c_k)}{1-q(c_k)}\right) - 	2q(c_k)\right) 
	\frac{1-q(c_k)}{2q(c_k)^2}}{\partial q(c_k)}\right|_{q(c_k) 
	= q^*}=0,
\end{equation*}
or taking the derivative and simplifying
\begin{equation*}
	(v-c_k) \frac{(q^{*2} - q^* - 2) \ln \left(\frac{1 - 
	q^*}{1 + q^*}\right)  - 2 q^* (2 + q^*)}{2 q^{*3} 
	(1 + q^*)}=0.
\end{equation*}
It is easy to see that $q^*$ is independent of $c_k$ which 
establishes fact (a).  To prove fact (b), we rewrite 
\eqref{eq:sk} as
\begin{equation}\label{eq:skbar}
	\overline{s}_k = (v-c_k)\left( \ln \left(\frac{1 + q^*_k}{1 
	- q^*_k}\right) - 2 q^*_k \right)\frac{1 - q^*_k}{2 
	q^{*2}_k}.
\end{equation}
It is easily observed that $\overline{s}_k$ is decreasing in 
$c_k$.  This completes the proof of part (i) of the corollary.

(ii)  Like in part (i), we start noting that 
\begin{equation}\label{eq:sbar}
	\begin{aligned}
\overline{s} &&=&&& \max_{q^U} \left\{\sum_{k=1}^K f_k 
\int_{0}^1 	\left(\frac{(1 - q^U)(v-c_k)}{1-q^U + 2q^U x} +	
c_k \right) (1-2x) dx\right\}\\
&&=&&& \max_{q^U} \left\{\sum_{k=1}^K f_k  (v-c_k)\left( \ln 	
\left(\frac{1 + q^U}{1 - q^U}\right) - 2 q^U\right) 
\frac{1-q^U}{2 (q^U)^2}\right\}.
	\end{aligned}
\end{equation}
Following a line of argument similar to that in part (i), one 
can establish that there exists a unique value of $q^U \in 
(0,1)$ that maximizes that expression in the curly brackets 
(noting that the expression in the curly brackets is 
concave in $q^U \in (0,1)$) and this unique value is independent 
of $c_k$s. It then follows that this unique value of $q^U$ is 
equal to $q^*$.  This means that $\overline{s}$ is the weighted 
average of $s_k$s.  Part (ii) of the corollary is complete.

\subsection{Proof of Proposition \ref{prop:welfare}}

To prove (i) we will first show that $q(c_k)$ is decreasing cna 
concave in $c_k$.  Second, letting $\overline{c} = \sum_{k=1}^K 
f_k c_k$ and noting that $q(\overline{c}) = q^U$, we will employ 
Jensen's inequality to demonstrate that show that $\sum_{k=1}^K 
f_k q(c_k)> q(\overline{c})=q^U$.  

Letting
\begin{equation*}
	A := \left( \ln \left(\frac{1+q(c_k)}{1-q(c_k)}\right) - 
	2q(c_k)\right)\frac{1-q(c_k)}{2q(c_k)^2},
\end{equation*}
we observe that the equilibrium $q(c_k)$ solves $A (v-c_k)=s.$  
Next, noting that in equilibrium it must be that 
$d(v-c_k)A/dc_k=0$, we obtain
\begin{equation}\label{eq:dq/dc}
	\frac{d q(c_k)}{dc_k} = \frac{A}{(v-c_k) \dfrac{\partial 
			A}{\partial q(c_k)}},
\end{equation}
which is negative as $\partial A/\partial q(c_k)<0$ in a stable 
equilibrium.  Differentiation of the both sides of the equation 
by $c_k$ once again yields
\begin{equation}\label{eq:d^q/dc^2}
	\frac{d^2 q(c_k)}{dc_k^2} = \frac{\left(\dfrac{\partial 
	A}{\partial q(c_k)}\right)^2 \dfrac{d q(c_k)}{d c_k} 
	(v-c_k)- A \left( -\dfrac{\partial A}{\partial q(c_k)} + 
	(v-c_k) \dfrac{\partial^2 A}{\partial q(c_k)^2} \dfrac{d 
	q(c_k)}{d c_k} \right)}{\left[(v-c_k)\dfrac{\partial 
	A}{\partial q(c_k)}\right]^2}.
\end{equation}
This is negative if the numerator of the RHS is negative.  Note 
that the first term in the numerator is negative, as $d q(c_k)/d 
c_k$ is negative. The expression in the large brackets of the 
second term in the numerator is positive as $\partial 
A/\partial q(c_k)<0$ and $\partial^2 A/\partial q(c_k)^2<0$ 
which follows from proof of Proposition \ref{prop:eq_asym} that 
the expected benefit of searching the second firm is concave in 
the search intensity, $q(c_k)$.%
\footnote{Alternatively, we can prove that $\partial^2A/\partial 
q(c_k)^2<0$ by directly differentiating $A$ twice w.r.t. 
$q(c_k)$.  Then, the inequality to be proved is
\begin{equation*}
	\frac{(1+q(c_k))^2 (3-q(c_k)) (1-q(c_k)) \ln 
	\left(\frac{1+q(c_k)}{1-q(c_k)}\right) - 2q(c_k) (3+ 
	2q(c_k)- 3q(c_k)^2 - q(c_k)^3)}{q(c_k)^4 (1-q(c_k)) 
	(1+q(c_k))^2} < 0
\end{equation*} 
for $0<q(c_k)<1$.  The inequality holds if it numerator is 
negative, or
\begin{equation}\label{eq:ineq_d2A}
	\ln \left(\frac{1+q(c_k)}{1-q(c_k)}\right)  < \frac{2q(c_k) 
	(3+ 2q(c_k)-3q(c_k)^2 - q(c_k)^3)}{(1+q(c_k))^2 	
	(3-q(c_k)) (1-q(c_k))}.
\end{equation}
Since both sides of the inequality converge to zero as $q(c_k) 
\downarrow 0$ and go to infinity as $q(c_k) \uparrow 1$, the 
inequality holds for any $0<q(c_k)<1$ if the derivative of its 
LHS w.r.t. $q(c_k)$ is less positive than that of the RHS.  The 
respective derivatives are $2/(1-q(c_k)^2)$ and $2(9+ 3q(c_k) - 
14 q(c_k)^2 + 2 q(c_k)^3 - q(c_k)^4 + 5 q(c_k)^5)/(1+q(c_k))^3 
(3-q(c_k))^2(1-q(c_k))^2$. Noting that $0<2/(1-q(c_k)^2)$  for 
any $0<q(c_k)<1$, we observe that the former derivative is less 
positive than the latter for any $0<q(c_k)<1$ if $9+ 3q(c_k) - 
14 q(c_k)^2 + 2 q(c_k)^3 - q(c_k)^4 + 5 q(c_k)^5 > (1+q(c_k))^2 
(3-q(c_k))^2(1-q(c_k))$, which can be simplified as $2q(c_k)^3(2 
- 3q(c_k) + 3q(c_k)^2)>0$.  It is easily verified that the last 
inequality is true for any $0<q(c_k)<1$.  This shows that the 
derivative of the LHS \eqref{eq:ineq_d2A} w.r.t. $q(c_k)$ is 
less positive than that of the RHS.  This in turn, along with 
the above limit results,  proves that the inequality in 
\eqref{eq:ineq_d2A} is true, or that $\partial^2 A/\partial 
q(c_k)^2 <0$. } This means that the numerator is indeed 
negative, and therefore $d^2q(c_k)/dc_k^2<0$ meaning that 
$q(c_k)$ is concave in $c_k$.

Concavity of $q(c_k)$ implies that $\sum_{k=1}^{K} f_k q(c_k) < 
q(\overline{c})$.  However, as $q(\overline{c})$ is independent 
of actual realization of the production marginal cost and solves 
\begin{equation*}
	\left( \ln \left(\frac{1 + q(\overline{c})}{1 - 
	q(\overline{c})}\right) - 2 q(\overline{c})\right)\frac{1 - 
	q(\overline{c})}{2 q(\overline{c}) ^2} \sum_{k=1}^K(v-c_k)=s,
\end{equation*}
it must be that $q(\overline{c}) = q^U$.  It then follows that 
$\sum_{k=1}^{K} f_k q(c_k) < q(\overline{c}) = q^U$, which 
completes the proof of the case where $\overline{s}_K<s$.

To prove (ii) we will show that the \textit{ex-interim} 
equilibrium consumer surplus, when consumer observe the 
production marginal cost, is concave in $c_k$.  We will then 
continue by demonstrating that the hypothetical consumer surplus 
for the production marginal cost equal to $\overline{c}$ equal 
to that with the information asymmetry on the production 
marginal cost and greater than the \textit{ex-ante} equilibrium 
consumer surplus without the information asymmetry.

If consumers observe the production marginal cost, their 
equilibrium (\textit{ex-interim}) surplus is
\begin{equation*}
		\begin{aligned}
CS(c_k) &&=&&& v - \int_0^1 p_k(x)(1-q(c_k) + q(c_k) 2x)dx - 
q(c_k) s \\
&&=&&&  v - \int_0^1 p_k(x)dx \\
&&=&&&  \left[1 - \frac{1-q(c_k)}{2q(c_k)} \ln 
\left(\frac{1+q(c_k)}{1-q(c_k)}\right)\right] (v-c_k)\\
&&=&&&  B(v-c_k),
		\end{aligned}
\end{equation*}
where we used \eqref{eq:qU} to obtain the second equality, 
simple algebraic manipulations to obtain the third equality, and 
let $B:=1 - \frac{1-q(c_k)}{2q(c_k)} \ln 
\left(\frac{1+q(c_k)}{1-q(c_k)}\right)$ to obtain the last 
equality.  To show that this consumer surplus is concave in 
$c_k$ we differentiate it twice w.r.t. $c_k$:
\begin{equation*}
	\begin{aligned}
\frac{d^2 B(v-c_k)}{d c_k^2} & = \frac{d }{dc_k}\left( - B 
+  	\frac{\partial B}{\partial q(c_k)}(v-c_k) \frac{d 
q(c_k)}{d c_k}\right) = 
\frac{d }{dc_k}\left( - B +  \frac{\partial B}{\partial 
q(c_k)} \times \frac{A}{\frac{\partial A}{\partial 
q(c_k)}}\right)\\
& = \frac{\partial }{ \partial q(c_k)}\left( - B +  
\frac{\partial B}{\partial q(c_k)} \times 
\frac{A}{\frac{\partial A}{\partial q(c_k)}}\right) \frac{d 
q(c_k)}{d c_k}\\
& = \left( - \frac{\partial B}{\partial q(c_k)} + 
\frac{\left(\frac{\partial^2 B}{\partial q^2(c_k)} A + 
\frac{\partial B}{\partial q(c_k)} \frac{\partial A}{\partial 
q(c_k)} \right)\frac{\partial A}{\partial q(c_k)} - 
\frac{\partial B}{\partial q(c_k)} \frac{\partial^2 A}{\partial 
q^2(c_k)} A }{\left(\frac{\partial A}{\partial 
q(c_k)}\right)^2}\right) \frac{d q(c_k)}{d c_k}\\
& = A \frac{\frac{\partial^2 B}{\partial q^2(c_k)} 
\frac{\partial A}{\partial q(c_k)} - \frac{\partial B}{\partial 
q(c_k)} \frac{\partial^2 A}{\partial q^2(c_k)} 
}{\left(\frac{\partial A}{\partial q(c_k)}\right)^2} \times 
\frac{d q(c_k)}{d c_k}.
	\end{aligned}
\end{equation*}
where we used \eqref{eq:dq/dc} to obtain the second equality in 
the first line.  Since $d q(c_k)/d c_k<0$, this double 
derivative is negative if 
\begin{equation}\label{eq:dAB}
	\frac{\partial^2 B}{\partial q^2(c_k)} 
	\frac{\partial A}{\partial q(c_k)} - \frac{\partial 
	B}{\partial q(c_k)} \frac{\partial^2 A}{\partial q^2(c_k)} > 
	0.
\end{equation}
As
\begin{equation*}
\Resize{}{
	\begin{aligned}
\frac{\partial A}{\partial q(c_k)} &= 
\frac{2 q(c_k) (2 + q(c_k)) - \left(2 + q(c_k) - q^2(c_k)\right) 
\ln \left(\frac{1 + q(c_k)}{1-q(c_k)}\right)}{2 q^3(c_k) (1 + 
q(c_k))},\\
\frac{\partial^2 A}{\partial q(c_k)^2} &=\frac{(1+q(c_k))^2 
(3-q(c_k)) (1-q(c_k)) \ln \left(\frac{1+q(c_k)}{1-q(c_k)}\right) 
- 2q(c_k) (3+ 2q(c_k)- 3q(c_k)^2 - q(c_k)^3)}{q(c_k)^4 
(1-q(c_k)) (1+q(c_k))^2},\\
\frac{\partial B}{\partial q(c_k)} &= \frac{(1+q(c_k))\ln 
\left(\frac{1+q(c_k)}{1-q(c_k)}\right) - 2 q(c_k)}{2 q(c_k)^2 
(1+q(c_k))},\\
\frac{\partial^2 B}{\partial q(c_k)^2} &= \frac{\frac{2 q(c_k) 
\left(1 + q(c_k) - q(c_k)^2\right)}{(1-q(c_k)) (1+q(c_k))^2} - 
\ln \left(\frac{1+ q(c_k)}{1-q(c_k)}\right)}{q(c_k)^3},
	\end{aligned}
}
\end{equation*}
the inequality can be rewritten (after some algebraic 
manipulations) as
\begin{equation*}
	\Resize{}{
	\dfrac{- 4 (1-2 q(c_k)) q(c_k)^2 + 2 q(c_k) \left(2 - q(c_k) 
	- q(c_k)^2\right) \ln \left(\frac{1+q(c_k)}{1-q(c_k)}\right) 
	- (1-q(c_k)) (1+q(c_k))^2 \left[\ln 
	\left(\frac{1+q(c_k)}{1-q(c_k)}\right)\right]^2}{2 q(c_k)^6 
	(1-q(c_k))  (1 +q(c_k))^2} >0.
}
\end{equation*}
As the denominator of the LHS is positive for any $0<q(c_k)<1$,  
the inequality is true if the numerator of the LHS is negative.  
As the numerator is a quadratic expression of 
$\ln((1-q)/(1+q))$, we verify that the numerator is negative if 
the following system of inequalities is true:
\begin{equation*}
	\begin{cases}
\ln \left(\dfrac{1+q(c_k)}{1-q(c_k)}\right) > 
\dfrac{2 q(c_k) + q(c_k)^2}{(1+q(c_k))^2} -  
\dfrac{q(c_k)^2}{(1+q(c_k))^2} 
\sqrt{\dfrac{9+7q(c_k)}{1-q(c_k)}},\\
\ln \left(\dfrac{1+q(c_k)}{1-q(c_k)}\right)  <  
\dfrac{2 q(c_k) + q(c_k)^2}{(1+q(c_k))^2} +  
\dfrac{q(c_k)^2}{(1+q(c_k))^2} 
\sqrt{\dfrac{9+7q(c_k)}{1-q(c_k)}}.
	\end{cases}
\end{equation*}

As $\sqrt{(9 + 7q(c_k))/(1-q(c_k))}>1$, the first inequality is 
certainly true if $\ln \left(\frac{1+q(c_k)}{1-q(c_k)}\right) > 
\frac{2 q(c_k)}{(1+q(c_k))^2}$. This inequality holds as, first, 
both sides of this inequality converge to zero as $q(c_k) 
\downarrow 0$; second, its LHS goes to infinity and RHS 
approaches $1/2$ as $q(c_k) \uparrow 1$; and third, the 
derivative of the LHS, which is $\frac{2}{1-q(c_k)^2}(>0)$, is 
greater than $2$ and that of the RHS, which is 
$\frac{2(1-q(c_k))}{(1+q(c_k))^3}(>0)$, is lower than $2$ for 
any $0<q(c_k)<1$. 

Letting $z := (1-q(c_k))/(1+q(c_k))$ where $z>1$ for 
$0<q(c_k)<1$, we rewrite the second inequality in the above 
system of inequalities as
\begin{equation*}
	\ln(z) < \frac{3z^2 - 2z - 1 + (z-1)^2 \sqrt{1+8z}}{4z^2}, \ 
	\ \mbox{for} \ \ z>1.
\end{equation*}
We prove the inequality with the help of the following three 
facts. First, both sides of the inequality are continuous in 
$z>1$.  Second, both sides of the inequality converge to zero as 
$z \downarrow 1$ and go to infinity as $z \to \infty$.  Finally, 
the LHS of the inequality is increasing more slowly than the RHS 
as the derivative of the LHS, which is $1/z (>0 \ \mbox{for} \ 
z>1)$, is smaller than that of the RHS, which is 
$\frac{(z-1)(1+6z + 2z^2) + (z+1) 
\sqrt{1+8z}}{2z^3\sqrt{1+8z}} (>0 \ \mbox{for} \ z>1)$:
\begin{equation*}
	\begin{aligned}
\sqrt{1+8z} (2z+1)(z-1) &&<&&& (z-1)(1+6z + 2z^2),\\
\sqrt{1+8z} (2z+1) &&<&&& (1+6z + 2z^2),\ \ |\ \mbox{square both 
sides to obtain},\\
1 + z2z + 36z^2 + 32 z^3 &&<&&& 1 + 12z + 40z^2 + 24z^3 + 4z^4,\\
0 &&<&&& 4z^2(z-1)^2.
	\end{aligned}
\end{equation*}

The arguments in the previous two paragraphs establish that the 
system of the inequality holds for any $0<q(c_k)<1$.  This, in 
turn, proves that the inequality in \eqref{eq:dAB} is true for 
any $0<q(c_k)<1$, which means that the equilibrium 
(\textit{ex-interim}) consumer surplus when they observe 
$c_k$---namely $B(v-c_k)$---is concave in $c_k$.

Concavity of the \textit{ex-interim} equilibrium consumer 
surplus when they observe the production marginal cost means 
that 
\begin{equation*}
	\begin{aligned}
	\sum_{k=1}^K f_k CS(c_k) &&=&&&  \sum_{k=1}^K f_k \left[1 - 
	\frac{1-q(c_k)}{2q(c_k)} \ln 
	\left(\frac{1+q(c_k)}{1-q(c_k)}\right)\right] (v-c_k) \\
	&& < &&& \left[1 - 
	\frac{1-q(\overline{c})}{2q(\overline{c})} \ln 
	\left(\frac{1+q(\overline{c})}{1-q(\overline{c})}\right) 
	\right] (v-\overline{c}) = CS(\overline{c}).
	\end{aligned}
\end{equation*} 
However, we know that $q(\overline{c}) = q^U$ and therefore 
$CS(\overline{c})$ equals to the consumer surplus when consumers 
do not observe the production marginal cost. This completes the 
proof of part (ii).

To prove (iii) we first show that the equilibrium 
\textit{ex-interim} profit, when consumers observe the marginal 
cost of production, is increasing and convex in $c_k$.  This 
implies that the corresponding equilibrium \textit{ex-ante} 
profit is lower than the profit given the expected production 
marginal cost, which coincides with equilibrium \textit{ex-ante} 
profit where consumers do not observe the production marginal 
cost. 

Let $\pi(c_k)$ be the equilibrium profit when $c_k$ is 
realized and buyers observe this realization, and so let $\Pi^O 
:= \sum_{k=1}^k f_k \pi(c_k)$ be the expected equilibrium 
profit.  We will show the following facts: $d \pi(c_k)/d 
c_k >0$ and $d^2 \pi(c_k)/dc_k^2>0$.  As $\pi(c_k) = 
(1-q(c_k))(v-c_k)$, we have that
\begin{equation*}
	\frac{d \pi(c_k)}{d c_k} = - (1-q(c_k)) - (v-c_k) \frac{d 
	q(c_k)}{d c_k} = - (1-q(c_k)) -  \frac{A}{\dfrac{\partial 
	A}{\partial q(c_k)}},
\end{equation*}
which is positive if $-(1-q(c_k)) \partial A/\partial q(c_k) - A 
< 0$ as $\partial A/\partial q(c_k)<0$. Substituting the values 
of $d A/dq(c_k)$ and $A$ and applying simple algebraic 
manipulations, we rewrite the inequality as 
$\ln((1+q(c_k))/(1-q(c_k)))< q(c_k)(2-q(c_k)^2)/(1-q(c_k)^2)$.  
Since both sides of the inequality converge to zero as $q(c_k) 
\downarrow 0$ and go to infinity as $q(c_k) \uparrow 1$, the 
inequality holds for any $0<q(c_k)<1$ if the derivative of the 
LHS w.r.t. $q(c_k)$ is less positive than that of the RHS. These 
derivatives are $2/(1-q(c_k)^2)$ and $(2-q(c_k)^2 + 
q(c_k)^4)/(1-q(c_k)^2)^2$, respectively; both are strictly 
positive for $0<q(c_k)<1$; and the former is smaller than the 
latter as $2 <(2-q(c_k)^2+q(c_k)^4)/(1-q(c_k)^2)$, which is 
easily verified to be true.  This proves that the derivative of 
the LHS of our initial inequality is less positive than that of 
the RHS, which in turn proves that our initial inequality holds. 
This means that $d \pi(c_k)/dc_k>0$.

We next note that 
\begin{equation*}
	\begin{aligned}
\frac{d^2 \pi(c_k)}{d c_k^2} &= \frac{d}{d c_k} \left(- 
(1-q(c_k)) -  \frac{A}{\dfrac{\partial A}{\partial 
q(c_k)}}\right) = \frac{\partial}{\partial q(c_k)} \left(- 
(1-q(c_k)) -  \frac{A}{\dfrac{\partial A}{\partial 
q(c_k)}}\right) \frac{d q(c_k)}{d c_k}\\
& = \left(1 - \frac{\left(\frac{\partial A}{\partial 
q(c_k)}\right)^2 - A \frac{\partial^2 A}{\partial 
q(c_k)^2}}{\left(\frac{\partial A}{\partial 
q(c_k)}\right)^2}\right) \frac{d q(c_k)}{d c_k} = \frac{ A 
\frac{\partial^2 A}{\partial q(c_k)^2}}{\left(\frac{\partial 
A}{\partial q(c_k)}\right)^2} \times 
\frac{d q(c_k)}{d c_k},
\end{aligned}
\end{equation*}
which is positive as $\partial^2 A/\partial q(c_k)^2 <0$ for 
$0<q(c_k)<1$, which follows from the proof of Proposition 
\ref{prop:eq_asym}. It is then indeed that $d^2 \pi(c_k)/d 
c_k^2>0$, or $\pi(c_k)$ is convex in $c_k$.

Convexity of $\pi(c_k)$ w.r.t. $c_k$ implies that for any 
non-degenerate $f$, it must be that $\sum_{k=1}^K 
f_k \pi(c_k) > \pi (\overline{c})$, or  
\begin{equation*}
	\sum_{k=1}^K f_k (1-q(c_k)(c_k))(v-c_k) > 
	(1-q(\overline{c}))(v- \overline{c}).
\end{equation*}
However, we know that $q(\overline{c}) = q^U$ and thus $\pi 
(\overline{c}) = (1-q(\overline{c}))(v- \overline{c}) = 
(1-q^U)(v- \overline{c}) = \Pi^U$, which is why $\Pi^O = 
\sum_{k=1}^K f_k \pi(c_k) > \Pi^U$.  This completes the proof of 
the proposition for $s \leq \overline{s}_K$.

\section{Additional Analysis}

\subsection{Truly Costly Search}\label{aa:trulycost}

To prove existence of a unique stable equilibrium with active 
search, it suffices to demonstrate that in equilibrium, 
searching yields a non-negative payoff.  We first consider a 
case where consumers observe the production marginal cost.  We 
then consider a case where consumers do not observe the 
production marginal cost.

Suppose consumers observe the production marginal cost.  We then 
have the following result which is reminiscent to Corollary 
\ref{cor:eq_sym}.

\begin{corollary}\label{cor:eq_sym_trulycost}
	Suppose buyers observe the production marginal cost. 
	\begin{itemize}
		\item[(i)]  There always exists the Diamond-paradox 
		equilibrium.
		\item[(ii)] For each $k = \{1,...,K\}$, there exists 
		$\overline{s}_k>0$ such that for $s\leq \overline{s}_k$ 
		there exists a unique stable NEs with active search 
		given in Corollary \ref{cor:eq_sym}.
	\end{itemize}
\end{corollary}

\begin{proof}
	
To prove the corollary, it suffices to show that for in a unique 
stable NE with active search, searching one firm yields a 
positive expected payoff. For any $c_k$, we know from the proof 
of Proposition \ref{prop:welfare} that searching one firm (or 
two firms) yields an expected payoff equal to
\begin{equation*}
	\left[1 - \frac{1-q(c_k)}{2q(c_k)} \ln 
	\left(\frac{1+q(c_k)}{1-q(c_k)}\right)\right] (v-c_k) - s.
\end{equation*}
As the expected price is decreasing in $q(c_k)$ and the search 
intensity is decreasing in $s$, the above expected payoff is not 
lower than
\begin{equation*}
	\left[1 - \frac{1-q^*}{2q^*} \ln 
	\left(\frac{1+q^*}{1-q^*}\right)\right] (v-c_k) - 
	\overline{s}_k,
\end{equation*}
where $q^*$ is the lowest possible search intensity in 
equilibrium with active search (recall the proof of Corollary 
\ref{cor:sk}). Using \eqref{eq:skbar}, we rewrite the above 
expression representing the lowest expected payoff (after 
applying simple algebraic manipulations) as
\begin{equation*}
	\left(1 + 2 \eta - 2 \eta \ln\left(1 + \frac{1}{\eta}\right) 
	- 2\eta^2 \ln\left(1 + \frac{1}{\eta}\right) \right)(v-c_k),
\end{equation*}
where $\eta := (1-q^*)/(2q^*)$ and so $\eta>0$ for any 
$0<q^*<1$.  This lowest payoff is positive if 
\begin{equation}\label{eq:ineq_eta}
	\frac{1+2\eta}{2\eta(1+\eta)} - \ln\left(1 + 
	\frac{1}{\eta}\right) \geq 0.
\end{equation}
We note the following facts about the LHS of the inequality. 
First, it is easily verified that the LHS is continuous in 
$\eta>0$.  Second, as 
$\lim_{\eta \to \infty} \frac{1+2\eta}{2(1+\eta)} =1$ and 
$\lim_{\eta \to \infty} \eta \ln\left(1 + \frac{1}{\eta}\right) 
= \lim_{z \downarrow 0} \frac{\ln(1+z)}{z} = \lim_{z \downarrow 
	0} \frac{1}{1+z}=1$, it follows that
\begin{equation*}
	\lim_{\eta \to \infty} \left[\frac{1+2\eta}{2\eta(1+\eta)} - 
	\ln\left(1 + \frac{1}{\eta}\right)\right] =  \lim_{\eta 
		\to \infty} \frac{\frac{1+2\eta}{2(1+\eta)}  - \eta 
		\ln\left(1 
		+ \frac{1}{\eta}\right)}{\eta} = 0.
\end{equation*}
Third, the derivative of the LHS of \eqref{eq:ineq_eta} w.r.t. 
$\eta$ is $-1/(2\eta^2(1+\eta)^2)$ which is negative for any 
$\eta > 0$. These three facts establish that the LHS of 
\eqref{eq:ineq_eta} is decreasing in $\eta$ and converges to 
zero as $\eta \to \infty$.  Therefore, the LHS of 
\eqref{eq:ineq_eta} must be positive for any $\eta>0$, i.e., the 
inequality is true.  The inequality implies that the lowest 
possible payoff in a stable equilibrium with active search must 
be positive.  This means that consumers prefer searching to not 
searching in a stable equilibrium with active search.  

The proof of the corollary is now complete.
\end{proof}

Suppose now that consumers do not observe the production 
marginal cost.  We then have the following result. 

\begin{corollary}\label{cor:eq_asym_trulycost}
	Suppose buyers do not observe the production marginal cost. 
	\begin{itemize}
		\item[(i)]  There always exists the Diamond-paradox 
		equilibrium.
		\item[(ii)] There exists $\overline{s}>0,$  such that 
		for $s< \overline{s}$ there exist a unique stable BNE 
		given in Proposition \ref{prop:eq_asym}.
	\end{itemize}
\end{corollary}

\begin{proof}

The proof follows a line of argument similar to the case where 
consumers observe the marginal cost of production.  Specifically 
in a stable equilibrium with active search, a searching consumer 
obtains an expected payoff not lower than
\begin{equation*}
	\begin{aligned}
		&&&\left[1 - \frac{1-q^*}{2q^*} \ln 
		\left(\frac{1+q^*}{1-q^*}\right)\right] \sum_{k=1}^K 
		f_k 	
		(v-c_k) - \overline{s} \\
		= &&&	\left(1 + 2 \eta - 2 \eta \ln\left(1 + 
		\frac{1}{\eta}\right) - 2\eta^2 \ln\left(1 + 
		\frac{1}{\eta}\right) \right)\sum_{k=1}^K f_k (v-c_k),
	\end{aligned}
\end{equation*} 
where we obtained the equality by using \ref{eq:sbar}.  We know 
that this payoff is positive from the case with the information 
asymmetry.  This proves that consumer payoff with search is 
positive in a stable equilibrium with active search.  

The proof of the corollary is now complete.

\end{proof}

\subsection{Price-comparing Consumers}\label{aa:shoppers}

In this subsection we provide full analysis of the model 
presented in Section \ref{ss:compare}.  Recall that the only 
difference of that model from our main model is the presence of 
$\lambda>0$ share of price-comparing consumers, where $\lambda$ 
is small. We will call the rest of the consumers \textit{costly 
consumers} as they need to incur search costs to compare prices. 
Our analysis consists of two parts.  We first identify 
conditions under which stable equilibria where costly consumers 
actively search, namely equilibria with active search. We then 
show that the information asymmetry on the production marginal 
cost benefits costly consumers.

We start the equilibrium analysis with the case where consumers 
observe the production marginal cost. An equilibrium with active 
search is determined by two conditions.  One is that an 
individual firm must be indifferent of setting any price in the 
support of the equilibrium price distribution and prefer these 
price to ones outside the support.  Call it \textit{equal profit 
condition}. The other conditions is that 
an individual costly consumers must be indifferent between 
searching one firm and searching two firms.  Call it 
\textit{indifference condition}. If we let 
\begin{equation*}
	\mu(q(c_k)) := \frac{(1-\lambda) (1-q(c_k))}{2 
	\left[(1-\lambda) (1-q(c_k)) + \lambda\right]}
\end{equation*}
be the share of consumers who observe one price over the share 
of price-comparing consumers, the equal profit condition yields
\begin{equation}\label{eq:xk_heter}
	x_k(p) = \mu(q(c_k)) \left(\frac{v-c}{p-c}-1\right) \ 
	\mbox{with support} \ \left[\underline{p}_k, v\right],
\end{equation}
where $\underline{p}_k$ solves $x_k(\underline{p})=1$.  If we 
let 
\begin{equation*}
	p_k(x) := \frac{\mu(q(c_k))}{\mu(q(c_k)) + x}(v-c_k) + c_k
\end{equation*}
be the inverse function, the indifference condition yields
\begin{equation}\label{eq:qk_heter}
	 \int_0^1  p_k(x)(1-2x)dx =s.
\end{equation}

We are not ready to state the equilibrium result.

\begin{proposition}\label{prop:NE_heter}
	Suppose buyers observe the production marginal cost. For 
	each $k = \{1,...,K\}$ and sufficiently small search costs, 
	a stable NE with active search exists and is given by 
	$(x_k,q(c_k))$ which are determined by \eqref{eq:xk_heter} 
	and \eqref{eq:qk_heter}.
\end{proposition}

\begin{proof}
A stable NE with active search exists for small search costs, if 
the LHS of \eqref{eq:qk_heter}, which represents the added 
benefit of 	searching the second firm, is positive in 
$0<q(c_k)<1$ and converges to zero as $q(c_k) \uparrow 1$. 

To show the former fact we first note that the LHS of 
\eqref{eq:qk_heter} equals to
\begin{equation*}
	\mu(q(c_k)) \left(\left[1 + 2\mu(q(c_k))\right]\ln\left(1  + 
	\frac{1}{\mu(q(c_k))}\right)-2\right)(v-c_k).
\end{equation*}
This is positive in $q(c_k) \in (0,1)$ if the terms in the large 
brackets are positive for $\mu(c_k)>0$, or if $\ln \left(1  + 
\frac{1}{\mu(q(c_k))}\right) > \frac{2}{1+2\mu(q(c_k))}$. As for 
$\mu(q(c_k)) \downarrow 0$ this latter inequality clearly holds, 
while the LHS and the RHS both approach 0 as $\mu(q(c_k)) 
\rightarrow \infty $, this inequality holds for all $\mu(q(c_k)) 
>0$ if the derivative of the LHS is more negative than that of 
the RHS. The derivate of the LHS is $-\frac{1}{\mu(q(c_k)) 
(1+\mu(q(c_k)))}$, while the derivate of the RHS is $-%
\frac{2}{(1+2\mu(q(c_k)))^{2}}$. It is easy to see that the 
former derivative is smaller than the latter. This establishes 
that the inequality holds for any $\mu(q(c_k))>0$, or 
$0<q(c_k)<1$, which in turn means that the LHS of 
\eqref{eq:qk_heter} is positive for any $0<q(c_k)<1$.

To show that the LHS of \eqref{eq:qk_heter} converges to zero as 
$q(c_k) \uparrow 1$, it suffices to show that is converges to 
zero as $\mu(q(c_k)) \downarrow 0$. As 
\begin{equation*}
	\begin{aligned}
\lim_{\mu(q(c_k)) \downarrow 0} \mu(q(c_k)) \ln\left(1  + 
\frac{1}{\mu(q(c_k))}\right) = \lim_{z \to \infty} 
\frac{\ln(1+z)}{z} \ \ \  \myeq \ \ \ \lim_{z \to \infty} 
\frac{1}{1+z} = 0,
	\end{aligned}
\end{equation*}
we indeed have that
\begin{equation*}
	\begin{aligned}
&&&\lim_{\mu(q(c_k)) \downarrow 0} \mu(q(c_k)) \left(\left[1 + 
2\mu(q(c_k))\right]\ln\left(1  + 
\frac{1}{\mu(q(c_k))}\right)-2\right)\\
 = &&& \lim_{\mu(q(c_k)) \downarrow 0} \mu(q(c_k)) \ln\left(1  + 
 \frac{1}{\mu(q(c_k))}\right)  + 2 \lim_{\mu(q(c_k)) \downarrow 
 0} \left[\mu(q(c_k))\right]^2  \ln\left(1  + 
 \frac{1}{\mu(q(c_k))}\right) \\
 &&& - 2 \lim_{\mu(q(c_k)) \downarrow 0} \mu(q(c_k))\\
= &&& 0.
	\end{aligned}
\end{equation*}

The proof of the proposition is now complete.
\end{proof}

We continue our equilibrium analysis by considering the case 
where consumers do not observe the production marginal cost.  
Like in the case where consumers observe the production marginal 
cost, a stable equilibrium with active search is determined by 
firms' $K$ number of equal profit conditions---an equal profit 
condition for each realization of the production marginal 
cost---and consumers' indifference condition.  If we let 
$\mu(q^U):= (1-\lambda) (1-q^U)/ 
\left[2 (1-\lambda) (1-q^U) + 2\lambda\right]$, equal profit 
condition for $c_k$ renders
\begin{equation}\label{eq:xk_sym_heter}
	x_k(p) = \mu(q^U) \left(\frac{v-c}{p-c}-1\right) \ 
	\mbox{with support} \ \left[\underline{p}_k, v\right]
\end{equation}
and with $x_k(\underline{p}_k)=1$. Using the inverse function
\begin{equation*}
	p_k(x) := \frac{\mu(q^U)}{\mu(q^U) + x}(v-c_k) + c_k,
\end{equation*}
we note that the indifference condition renders
\begin{equation}\label{eq:qu_heter}
	\sum_{k=1}^{K}f_k \int_0^1  p_k(x)(1-2x)dx =s.
\end{equation}

We are now in a position to state the equilibrium result with 
the information asymmetry on the production marginal cost.
\begin{proposition}\label{prop:BNE_heter}
	Suppose buyers do not observe the production marginal cost. 
	For sufficiently small search costs, there exists a stable 
	BNE with active search given by $((x_k)_{k=1}^K,q^U)$ where 
	each $x_k$ is determined by \eqref{eq:xk_sym_heter} and 
	$q^U$ by \eqref{eq:qu_heter}.
\end{proposition}

\begin{proof}
The proof follows a line of argument similar to the proof of 	
Proposition \ref{prop:NE_heter}. Specifically it suffices to 	
show that the LHS of \eqref{eq:qu_heter} is positive for any 	
$\mu(q^U)>0$ and converges to zero as $\mu(q^U) \downarrow 0$. 

We first rewrite the LHS of \eqref{eq:qu_heter} as
\begin{equation*}
	\sum_{k=1}^{K} f_k \mu(q^U) \left(\left[1 + 
	2\mu(q^U)\right]\ln\left(1  + 
	\frac{1}{\mu(q^U)}\right)-2\right) (v-c_k).
\end{equation*}
However, We know from the proof of Proposition 
\ref{prop:NE_heter} that 
the terms under the summation sign are positive for any 
$\mu(q^U)>0$.  We also know that those terms converge to zero as 
$\mu(q^U) \downarrow 0$.   Thus the proof of the proposition is 
complete.
\end{proof}

We now turn our attention to the welfare analysis. The main 
result is stated in the following proposition.

\begin{proposition}\label{prop:welfare_heter}
	For sufficiently small search costs and the share of 
	price-comparing consumers, in equilibrium
	\begin{itemize}
		\item[(i)] consumers search less intensely,
		\item [(ii)] surplus of costly consumers is lower, 
		\item[(iii)] firm profit is higher, and
		\item[(iv)] total surplus is higher
	\end{itemize}
	in expectation when consumers observe the production 
	marginal cost than when they do not.
\end{proposition}

\begin{proof}
To prove the proposition it is useful to let $\phi:= 	
2\lambda/(1-\lambda)$ so that $\mu(q(c_k)) = (1-q(c_k))/(2 	
q(c_k) + \phi).$  We can then rewrite \eqref{eq:qk_heter} as	
\begin{equation*}
	\int_{0}^{1}\left(\frac{1-q(c_k)}{1-q(c_k) + \left(2 q(c_k) 
	+ \phi\right)x}(v-c) + c_k\right) (1-2x)dx=s.
\end{equation*}
Notice that this equation is continuous in $\lambda \in [0,1)$ 
and converges to \eqref{eq:qk} as $\lambda  \downarrow 0$, or 
$\phi \downarrow 0$.  It then follows that for sufficiently 
small $\lambda$, all our results in the main body of the paper 
must hold. This completes the proof of the proposition.
\end{proof}

\end{singlespace}

\newpage

\bibliographystyle{ecta}

\end{document}